\documentclass{article}

\usepackage[final,nonatbib]{nips_2018}

\def \hfillx {\hspace*{-\textwidth} \hfill}
\usepackage[utf8]{inputenc} 
\usepackage[T1]{fontenc}    
\usepackage{hyperref}       
\usepackage{url}            
\usepackage{booktabs}       
\usepackage{amsfonts}       
\usepackage{nicefrac}       
\usepackage{microtype}      
\usepackage{amsmath,graphicx}
\usepackage{multirow}
\usepackage{subcaption}
\usepackage{caption}
\usepackage{subcaption}
\usepackage{cool}

\title{Interpretable Convolutional Filters with SincNet}
\author{
  Mirco Ravanelli\\
  Mila, Universit\'e de Montr\'eal \\
  \And
  Yoshua Bengio \\
  Mila, Universit\'e de Montr\'eal \\
  CIFAR Fellow
}

\begin{document}

\maketitle

\begin{abstract}
Deep learning is currently playing a crucial role toward higher levels of artificial intelligence.
This paradigm allows neural networks to learn complex and abstract representations, that are progressively obtained by combining simpler ones.
Nevertheless, the internal "black-box" representations automatically discovered by current neural architectures often suffer from a lack of interpretability, making of primary interest the study of explainable machine learning techniques.

This paper summarizes our recent efforts to develop a more interpretable neural model for directly processing speech from the raw waveform. In particular, we propose \textit{SincNet}, a novel Convolutional Neural Network (CNN) that encourages the first layer to discover more meaningful filters by exploiting parametrized sinc functions. In contrast to standard CNNs, which learn all the elements of each filter, only low and high cutoff frequencies of band-pass filters are directly learned from data. This inductive bias offers a very compact way to derive a customized filter-bank front-end, that only depends on some parameters with a clear physical meaning. 
Our experiments, conducted on both speaker and speech recognition, show that the proposed architecture converges faster, performs better, and is more interpretable than standard CNNs.

\end{abstract}

\section{Introduction}
Deep learning has recently contributed to achieving unprecedented performance levels in numerous tasks, mainly thanks to the progressive maturation of supervised learning techniques \cite{Goodfellow-et-al-2016-Book}. 
The increased discrimination power of modern neural networks, however, is often obtained  at  the  cost of  a reduced interpretability  of the model. Modern end-to-end systems, whose popularity is increasing in many fields such as speech recognition \cite{lideng,attention_asr,graves_ctc}, often discover "black-box" internal representations that make sense for the machine but are arguably difficult to interpret by humans.
The remarkable sensitivity of current neural networks toward adversarial examples \cite{adversarial}, for instance, not only highlights how superficial the discovered representations could be but also raises crucial concerns about our capabilities to really interpret neural models.
Such a lack of interpretability can be a major bottleneck for the development of future deep learning techniques. Having more meaningful insights on the logic behind network predictions and errors, in fact, can help us to better trust, understand, and diagnose our model, eventually guiding our efforts toward more robust deep learning. 
In recent years, a growing interest has been thus devoted to the development of interpretable machine learning \cite{book_interp,int_summary}, as witnessed by the numerous works in the field, ranging from visualization \cite{understand_cnn,int_visual}, diagnosis of DNNs  \cite{lime}, explanatory graphs \cite{exp_graph}, and explainable models \cite{capsule}, just to name a few.

Interpretability is a major concern for audio and speech applications as well \cite{int_audio}. CNNs and Recurrent Neural Networks (RNNs) are the most popular architectures nowadays used in speech and speaker recognition \cite{lideng}. RNN can be employed to capture the temporal evolution of the speech signal \cite{lstm,gru2,ravanelli_is17,li_gru}, while CNNs, thanks to their weight sharing, local filters, and pooling  networks are normally employed to extract robust and invariant representations \cite{cnn_lecun}. Even though standard hand-crafted features such as FBANK and Mel-Frequency Cepstral Coefficients (MFCC)  are still employed in many state-of-the-art systems \cite{dnn_spk_rec_class2,dnn_speaker_rec_plp,spk_id_mfcc}, directly feeding a CNN with spectrogram bins \cite{e2e_spk_id,spk_rec_time_freq,voxceleb} or even with raw audio samples \cite{palaz_raw,tara_raw,google_rawmulti,joint7,tuske,dnn_emotion,wavenet,sample_rnn,acoustic_raw_povey,spoofing_raw,raw_speaker_id,verification_raw_ICASSP2018,verification_raw_IS2018} is an approach of increasing popularity.
The engineered features, in fact, are originally designed from perceptual evidence and there are no guarantees that such representations are optimal for all speech-related tasks. Standard features, for instance, smooth the speech spectrum, possibly hindering the extraction of crucial narrow-band speaker characteristics such as pitch and formants. Conversely, directly processing the raw waveform allows the network to learn low-level representations that are possibly more customized on each specific task.

The downside of raw speech processing lies in the possible lack of interpretability of the filter bank learned in the first convolutional layer. According to us, the latter layer is arguably the most critical part of current waveform-based CNNs. This layer deals with high-dimensional inputs and is also more affected by vanishing gradient problems, especially when employing very deep architectures. As will be discussed in this paper, the filters learned by CNNs often take noisy and incongruous multi-band shapes, especially when few training samples are available. These filters certainly make some sense for the neural network, but they do not appeal to human intuition, nor appear to lead to an efficient representation of the speech signal. 

To help the CNNs discover more meaningful filters, this work proposes to add some constraints on their shape. Compared to standard CNNs, where the filter-bank characteristics depend on several parameters (each element of the filter vector is directly learned), SincNet convolves the waveform with a set of parametrized sinc functions that implement band-pass filters \cite{SincNet}. The low and high cutoff frequencies are the only parameters of the filter learned from data. This solution still offers considerable flexibility but forces the network to focus on high-level tunable parameters that have a clear physical meaning. 
Our experimental validation has considered both speaker and speech recognition tasks. Speaker recognition is carried out on TIMIT \cite{timit} and Librispeech \cite{librispeech} datasets under challenging but realistic conditions, characterized by minimal training data (i.e., 12-15 seconds for each speaker) and short test sentences (lasting from 2 to 6 seconds). With the purpose of validating SincNet in both clean and noisy conditions, speech recognition experiments are conducted on both the TIMIT and DIRHA dataset \cite{dirha_asru,rav_is16}. 
Results show that the proposed SincNet converges faster, achieves better performance, and is more interpretable than a more standard CNN.

The remainder of the paper is organized as follows. The SincNet architecture is described in Sec.~\ref{sec:sinc}. Sec. \ref{sec:rel_work} discusses the relation to prior work. The experimental activity on both speaker and speech recognition is outlined in  Sec.~\ref{sec:exp}. Finally, Sec.~\ref{sec:conc} discusses our conclusions.

\section{The SincNet Architecture} \label{sec:sinc}
\begin{figure}[t!]
 \centering
   \includegraphics[scale=0.70]{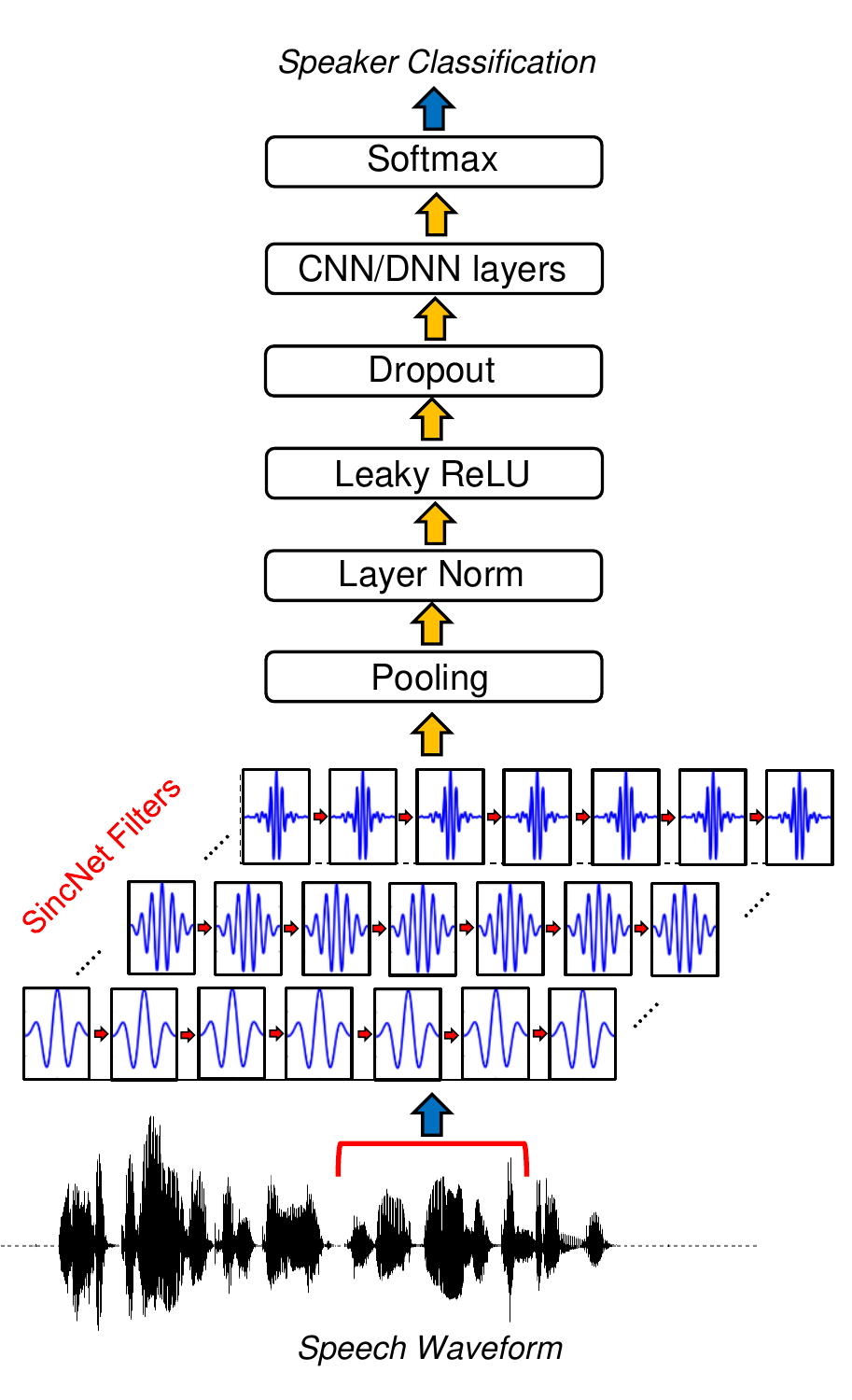}
 \caption{Architecture of SincNet.}
 \label{fig:sinc_arch}
 \end{figure}

The first layer of a standard CNN performs a set of time-domain convolutions between the input waveform and some Finite Impulse Response (FIR) filters \cite{rabiner11}. Each convolution is defined as follows\footnote{Most deep learning toolkits actually compute \textit{correlation} rather than \textit{convolution}. The obtained flipped (mirrored) filters do not affect the results.}:
\begin{equation}
y[n]=x[n]*h[n] = \sum\limits_{l=0}^{L-1} x[l]\cdot h[n-l] 
\end{equation}
where $x[n]$ is a chunk of the speech signal, $h[n]$ is the filter of length $L$, and $y[n]$ is the filtered output. In standard CNNs, all the L elements (taps) of each filter are learned from data. Conversely, the proposed SincNet (depicted in Fig. \ref{fig:sinc_arch}) performs the convolution with a predefined function $g$ that depends on few learnable parameters $\theta$ only, as highlighted in the following equation:

\begin{equation}
y[n]=x[n]*g[n,\theta] 
\end{equation}

A reasonable choice, inspired by standard filtering in digital signal processing, is to define $g$ such that a filter-bank composed of rectangular bandpass filters is employed. In the frequency domain, the magnitude of a generic bandpass filter can be written as the difference between two low-pass filters:

\begin{equation}
G[f,f_1,f_2]= rect\Big(\frac{f}{2f_{2}}\Big) - rect\Big(\frac{f}{2f_{1}}\Big),
\end{equation}
where $f_{1}$ and $f_{2}$ are the learned low and high cutoff frequencies, and $rect(\cdot)$ is the rectangular function in the magnitude frequency domain\footnote{The phase of the $rect(\cdot)$ function is considered to be linear.}.
After returning to the time domain (using the inverse Fourier transform \cite{rabiner11}), the reference function $g$ becomes:

\begin{equation}
g[n,f_1,f_2]= 2f_{2}sinc(2\pi f_2n) - 2f_{1}sinc(2\pi f_1n),
\end{equation}
where the sinc function is defined as $sinc(x)=sin(x)/x$. 

The cut-off frequencies can be initialized randomly in the range $[0,f_s/2]$, where $f_s$ represents the sampling frequency of the input signal.  
As an alternative, filters can be initialized with the cutoff frequencies of the mel-scale filter-bank, which has the advantage  of directly allocating more filters in the lower part of the spectrum, where crucial speech information is located.
To ensure $f_1\geq0$ and $f_2 \geq f_1$, the previous equation is actually fed by the following parameters:

\begin{align} 
&f_1^{abs}=|f_1| \\ 
&f_2^{abs}=f_1+|f_2-f_1|
\end{align}

Note that no bounds have been imposed to force $f_2$ to be smaller than the Nyquist frequency, since we observed that this constraint is naturally fulfilled during training. 
Moreover, the gain of each filter is not learned at this level. This parameter is managed by the subsequent layers, which can easily attribute more or less importance to each filter output.

An ideal bandpass filter (i.e., a filter where the passband is perfectly flat and the attenuation in the stopband is infinite) requires an infinite number of elements $L$. Any truncation of $g$ thus inevitably leads to an approximation of the ideal filter, characterized by ripples in the passband and limited  attenuation  in  the  stopband.   A popular solution to mitigate this issue is windowing \cite{rabiner11}. Windowing is performed by multiplying the truncated function $g$ with a window function $w$, which aims to smooth out the abrupt discontinuities at the  ends  of  $g$:
\begin{equation}
g_{w}[n,f_1,f_2]= g[n,f_1,f_2] \cdot w[n].
\end{equation}
This paper uses the popular Hamming window \cite{mitra}, defined as follows:
\begin{equation}
w[n]= 0.54-0.46 \cdot cos\Big(\frac{2\pi n}{L}\Big).
\end{equation}
The Hamming window is particularly suitable to achieve high frequency selectivity \cite{mitra}. However, results not reported here reveal no significant performance difference when adopting other functions, such as Hann, Blackman, and Kaiser windows. Note also that the filters $g$ are symmetric and thus do
not introduce any phase distortions. Due to the symmetry, the
filters can be computed efficiently by considering one side of
the filter and inheriting the results for the other half

All operations involved in SincNet are fully differentiable and the cutoff frequencies of the filters can be jointly optimized with other CNN parameters using Stochastic Gradient Descent (SGD) or other gradient-based optimization routines. 
As shown in Fig.  \ref{fig:sinc_arch}, a standard CNN pipeline (pooling, normalization, activations, dropout) can be employed after the first sinc-based convolution.
Multiple standard convolutional, fully-connected or recurrent layers \cite{gru2,ravanelli_is17,li_gru,ravanelli_twin} can then be stacked together to finally perform a  classification with a softmax classifier. 

\begin{figure*}[t!]
\begin{subfigure}{0.5\textwidth}
\includegraphics[scale=0.525,trim={0cm 0cm 0cm 0cm},clip]{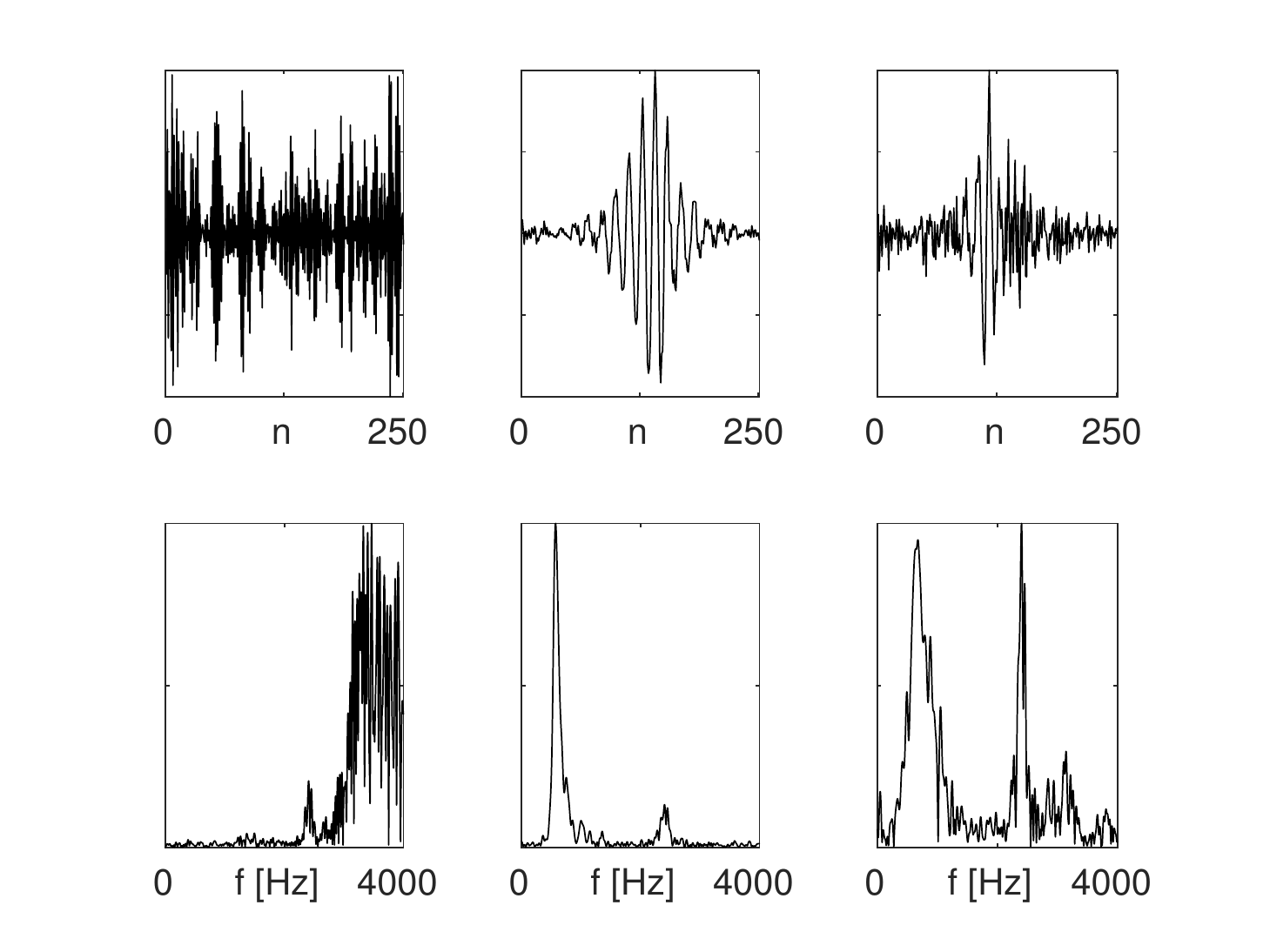}
\caption{CNN Filters}
\label{fig:cnn_filt}
\end{subfigure} \hspace{0.0\textwidth}
\begin{subfigure}{0.50\textwidth}
\includegraphics[scale=0.525]{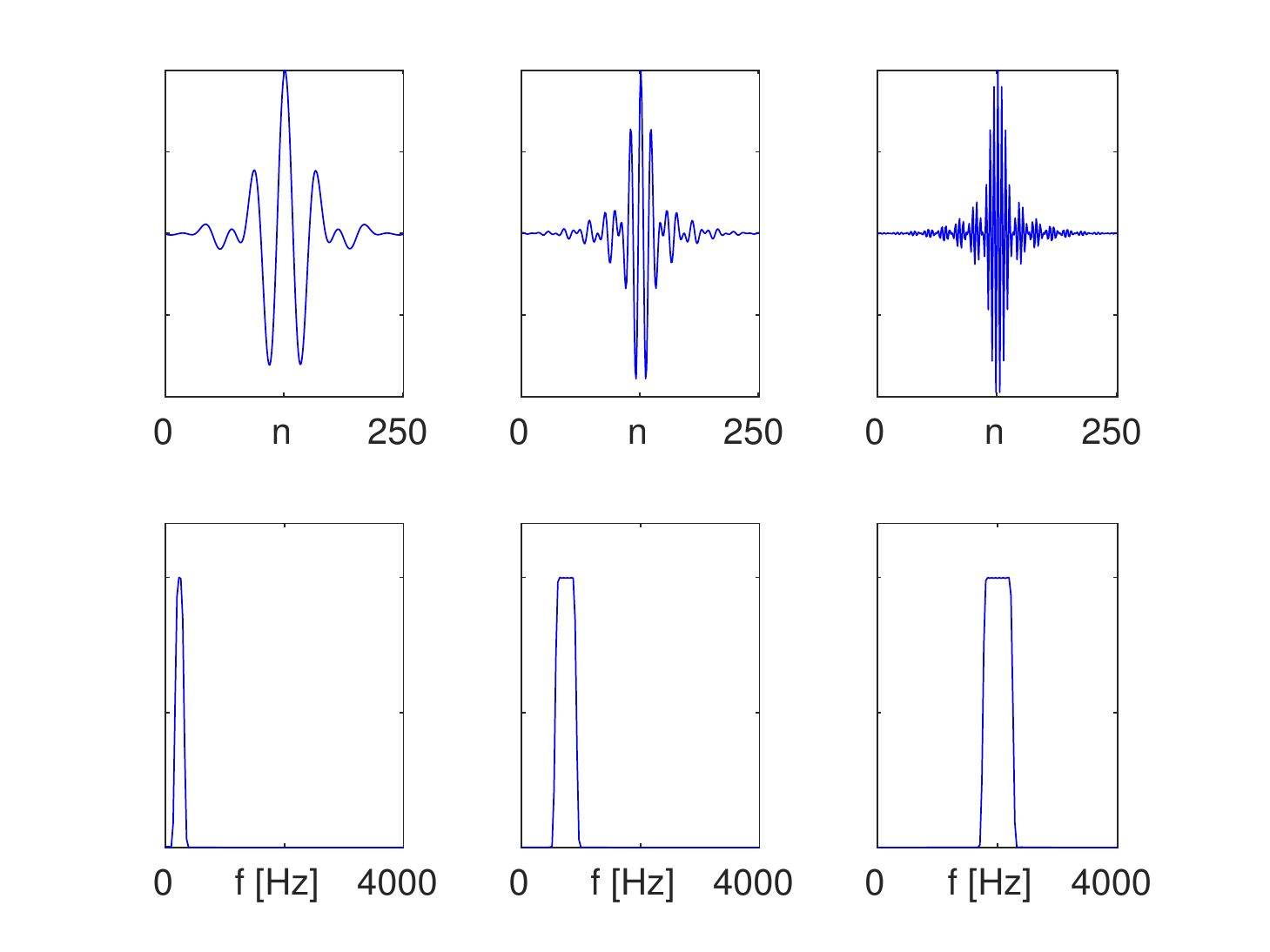}
\caption{SincNet Filters}
\label{fig:sinc_filt}
\end{subfigure}
\caption{Examples of filters learned by a standard CNN and by the proposed SincNet (using the Librispeech corpus on a speaker-id task). The first row reports the filters in the time domain, while the second one shows their magnitude frequency response.}
\label{fig:ir}
\end{figure*}

Fig. \ref{fig:ir} shows some examples of filters learned by a standard CNN and by the proposed SincNet for a speaker identification task trained on Librispeech (the frequency response is plotted between 0 and 4 kHz). As observed in the figures, the standard CNN does not always learn filters with a well-defined frequency response. In some cases, the frequency response looks noisy (see the first CNN filter), while in others assuming  multi-band shapes (see the third CNN filter). SincNet, instead, is specifically designed to implement rectangular bandpass filters, leading to more a meaningful filter-bank.

\subsection{Model properties}

\begin{figure}
\centering
\begin{minipage}{.5\textwidth}
  \centering
  \includegraphics[scale=.50,trim={0cm 0cm 0cm 0cm},clip]{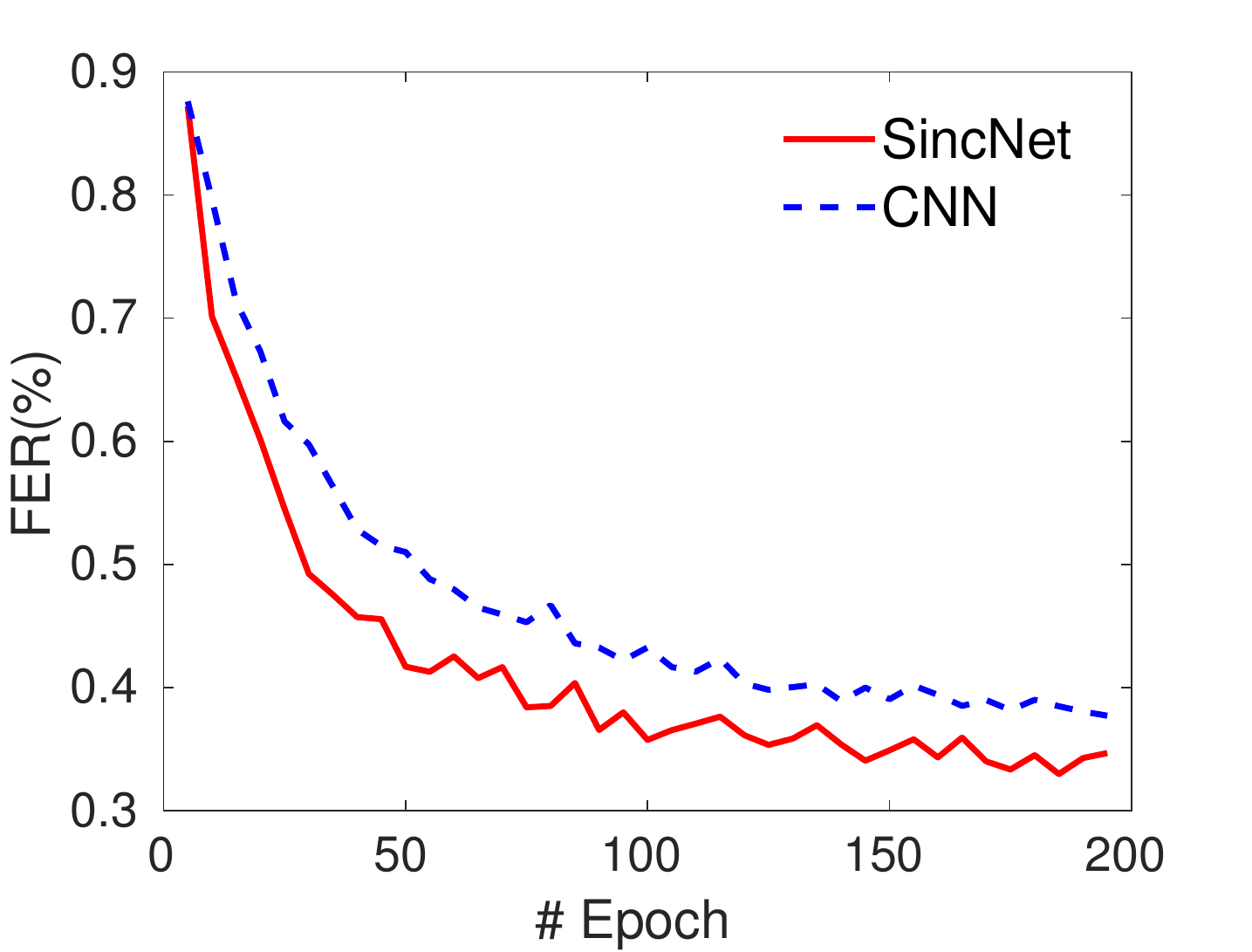}
  \captionsetup{width=.90\textwidth}
  \captionof{figure}{Frame Error Rate (\%) obtained on speaker-id with the TIMIT corpus (using held-out data).}
  \label{fig:conv_curve}
\end{minipage}%
\begin{minipage}{.5\textwidth}
  \centering
  \includegraphics[scale=.50]{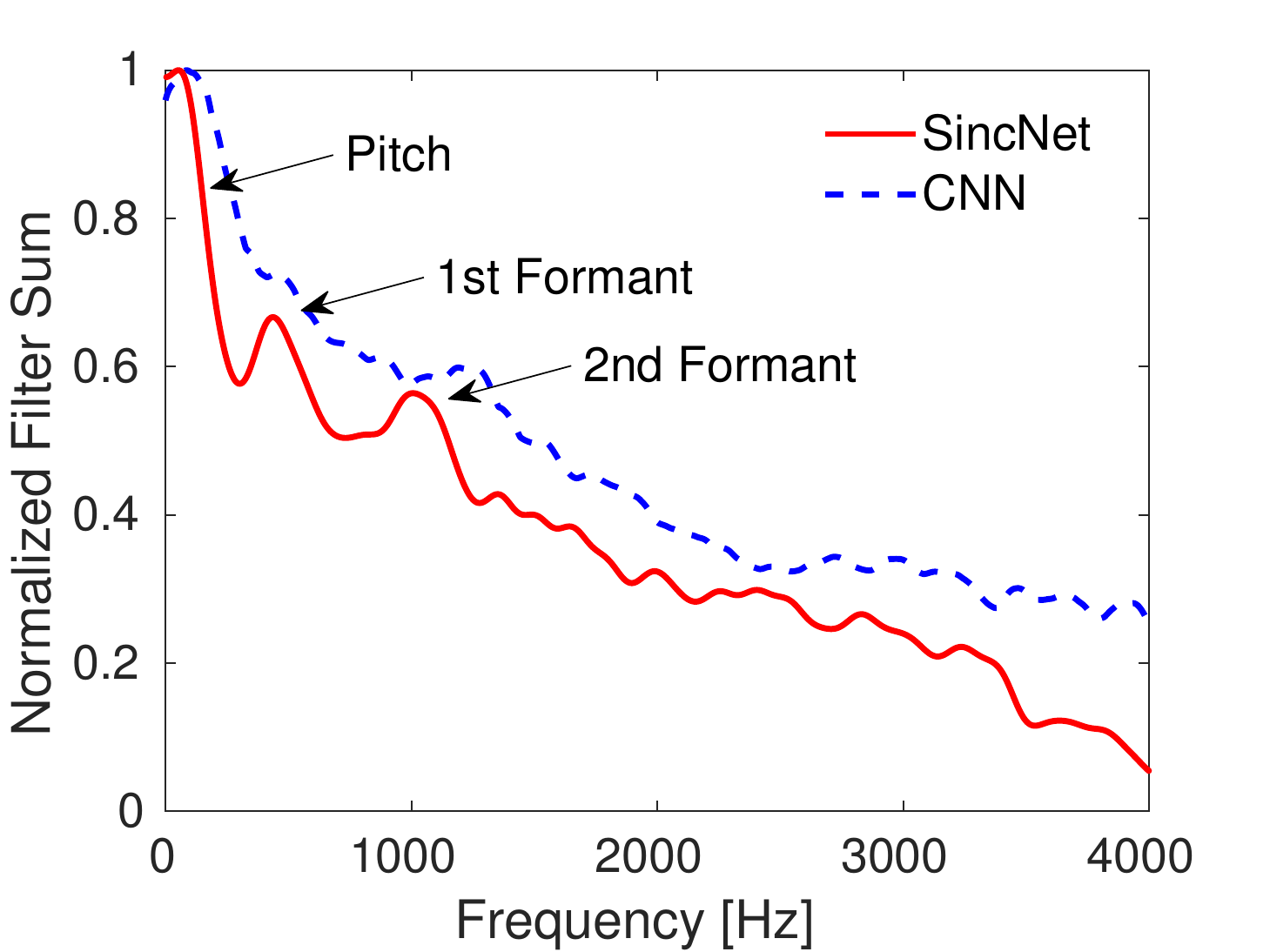}
  \captionsetup{width=.90\textwidth}
  \captionof{figure}{Cumulative frequency response of SincNet and CNN filters on speaker-id.}
  \label{fig:cum}
\end{minipage}
\end{figure}

The proposed SincNet has some remarkable properties:
\begin{itemize}
\item \textbf{Fast Convergence:}
SincNet forces the network to focus only on the filter parameters with a major impact on performance. The proposed approach actually implements a natural inductive bias, utilizing knowledge about the filter shape (similar to feature extraction methods generally deployed on this task) while retaining flexibility to adapt to data. This prior knowledge makes learning the filter characteristics much easier, helping SincNet to converge significantly faster to a better solution.
Fig. \ref{fig:conv_curve} shows the learning curves of SincNet and CNN obtained in a speaker-id task. These results are achieved on the TIMIT dataset and highlight a faster decrease of the Frame Error Rate ($FER\%$) when SincNet is used. Moreover, SincNet converges to better performance leading to a FER of 33.0\% against a FER of 37.7\% achieved with the CNN baseline.

\item \textbf{Few Parameters:} SincNet drastically reduces the number of parameters in the first convolutional layer. 
For instance, if we consider a layer composed of $F$ filters of length $L$, a standard CNN employs $F \cdot L$ parameters, against the $2F$ considered by SincNet. If $F=80$ and $L=100$, we employ 8k parameters for the CNN and only 160 for SincNet. Moreover, if we double the filter length $L$, a standard CNN doubles its parameter count (e.g., we go from 8k to 16k), while SincNet has an unchanged parameter count (only two parameters are employed for each filter, regardless its length $L$). This offers the possibility to derive very selective filters with many taps, without actually adding parameters to the optimization problem. Moreover, the compactness of the SincNet architecture makes it suitable in the few sample regime. 


\item \textbf{Interpretability}: The SincNet feature maps obtained in the first convolutional layer are definitely more interpretable and human-readable than other approaches. The filter bank, in fact, only depends on parameters with a clear physical meaning.
Fig. \ref{fig:cum}, for instance, shows the cumulative frequency response of the filters learned by SincNet and CNN on a speaker-id task. The cumulative frequency response is obtained by summing up all the discovered filters and is useful to highlight which frequency bands are covered by the learned filters. 

\begin{figure*}[t!]

\begin{subfigure}{0.5\textwidth}
\includegraphics[scale=0.850,trim={0cm 0cm 0cm 0cm},clip]{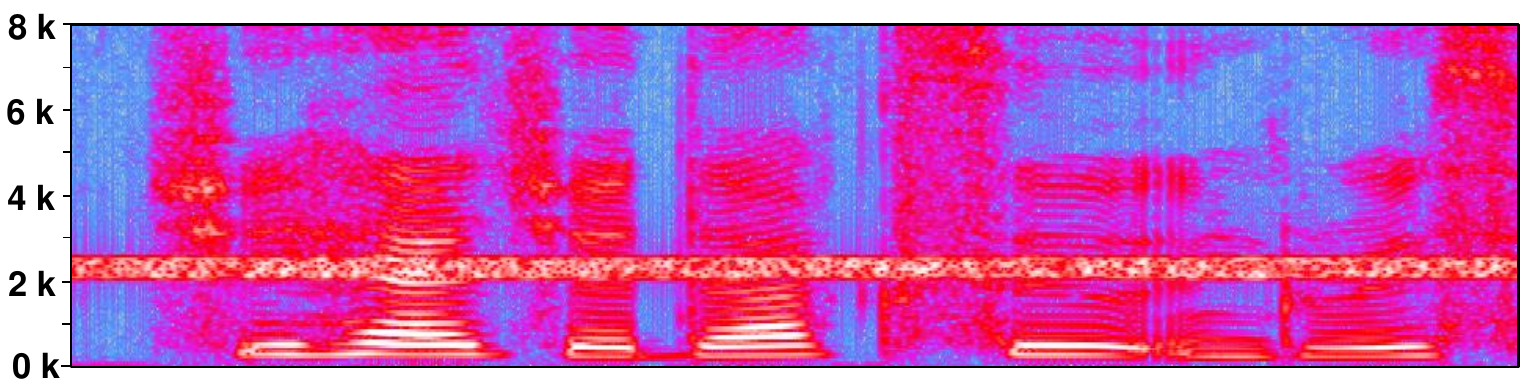}
\label{fig:cnn_filt}
\end{subfigure} \hspace{0.0\textwidth \vspace{0em}}

\begin{subfigure}{0.5\textwidth}
\includegraphics[scale=0.450,trim={0cm 0cm 0cm 0cm},clip]{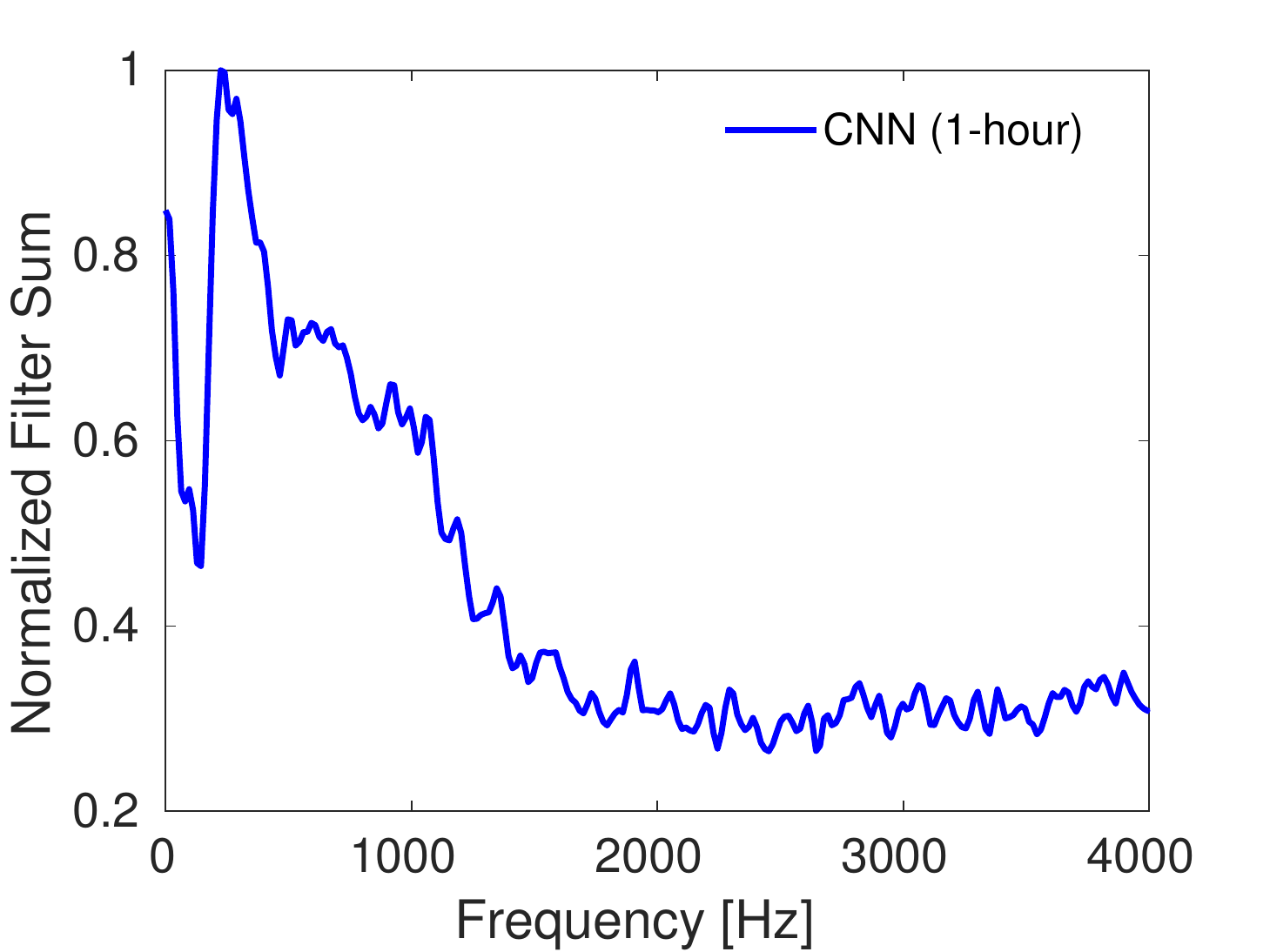}
\label{fig:cnn_filt}
\end{subfigure} \hspace{0.0\textwidth}
\begin{subfigure}{0.50\textwidth}
\includegraphics[scale=0.450]{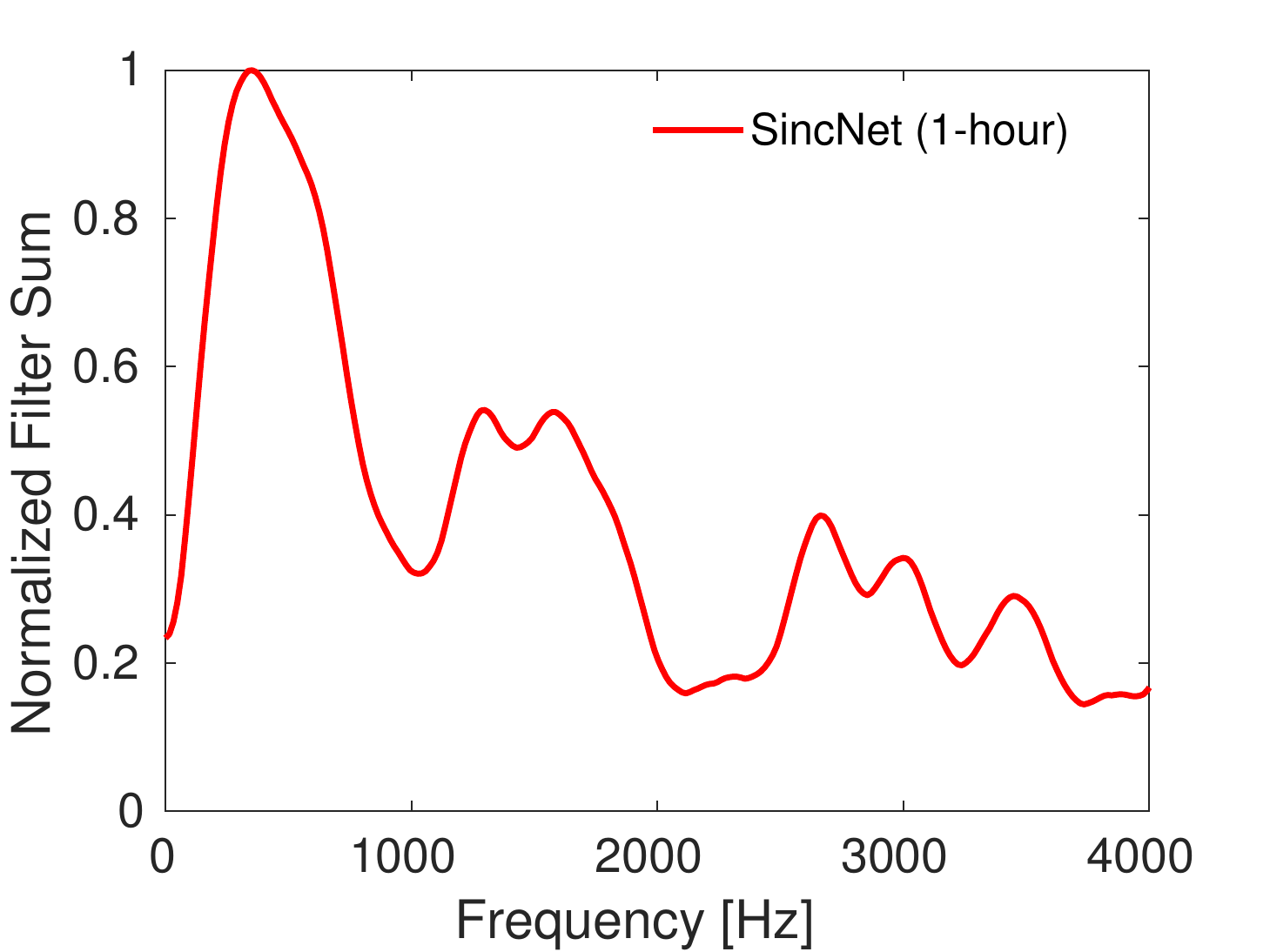}
\label{fig:sinc_filt}
\end{subfigure}

\begin{subfigure}{0.5\textwidth}
\includegraphics[scale=0.450,trim={0cm 0cm 0cm 0cm},clip]{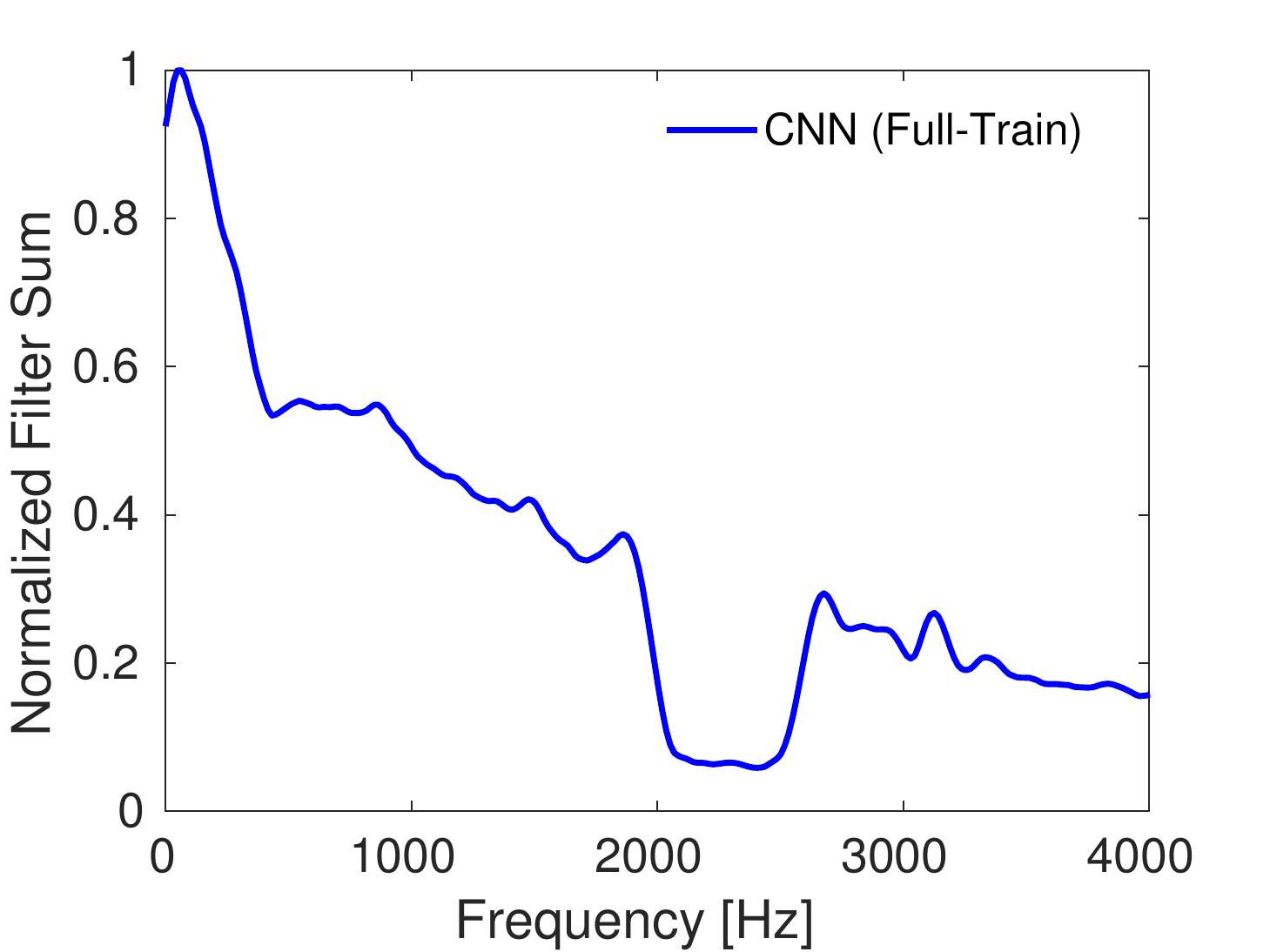}
\label{fig:cnn_filt}
\end{subfigure} \hspace{0.0\textwidth}
\begin{subfigure}{0.50\textwidth}
\includegraphics[scale=0.450]{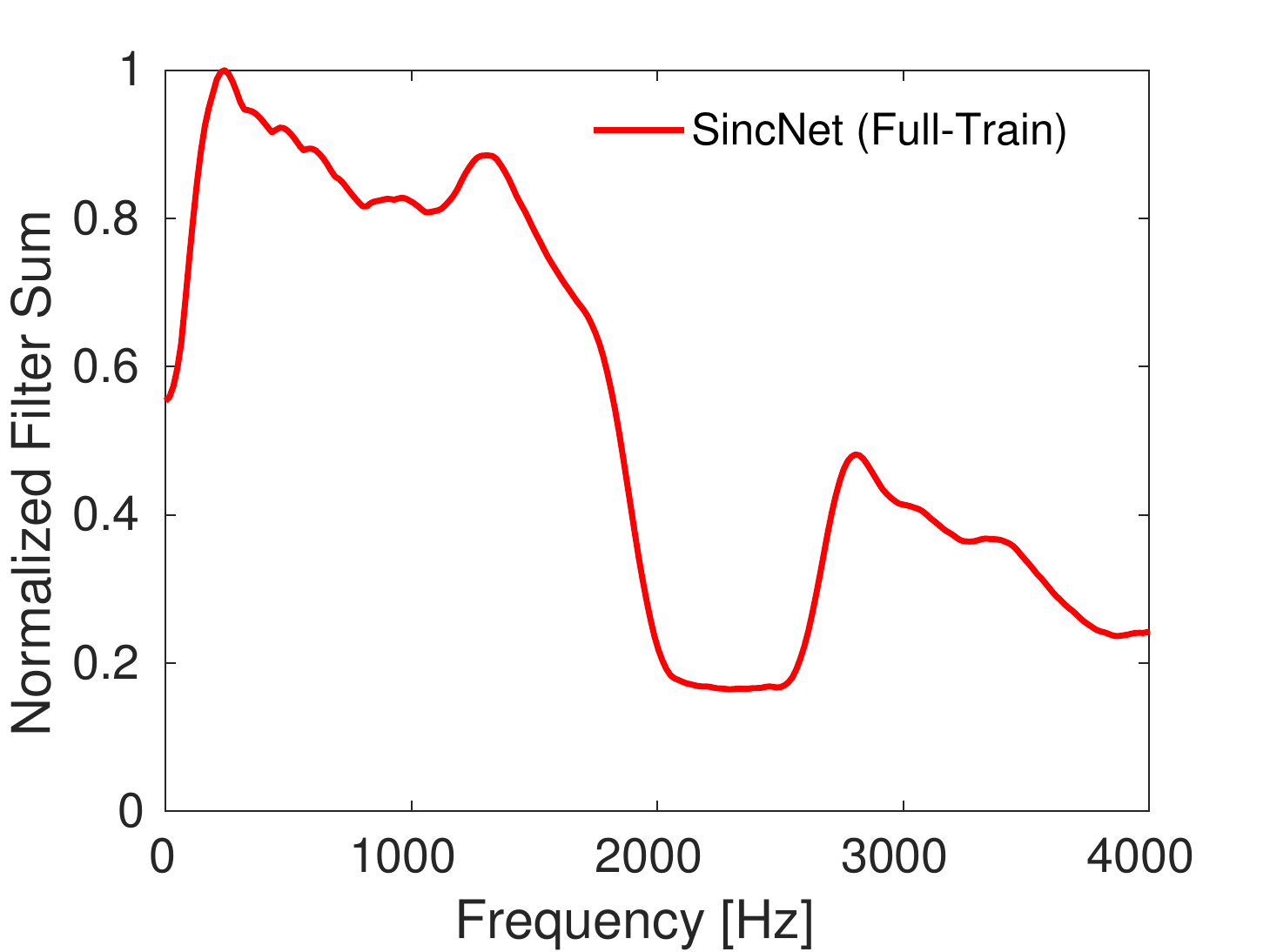}
\label{fig:sinc_filt}
\end{subfigure}

\caption{Cumulative frequency responses obtained on a speech recognition task trained with a noisy version of TIMIT. As shown in the spectrogram, noise has been artificially added into the band 2.0-2.5 kHz. Both the CNN and SincNet learn to avoid the noisy band, but SincNet learns it much faster, after processing only one hour of speech.}
\label{fig:asr_cum}
\end{figure*}

Interestingly, there are three main peaks which clearly stand out from the SincNet plot (see the red line in the figure). The first one corresponds to the pitch region (the average pitch is 133 Hz for a male and 234 for a female). The second peak (approximately located at 500 Hz) mainly captures first formants, whose average value over the various English vowels is indeed 500 Hz. Finally, the third peak (ranging from 900 to 1400 Hz) captures some important second formants, such as the second formant of the vowel $/a/$, which is located on average at 1100 Hz.
This filter-bank configuration indicates that SincNet has successfully adapted its characteristics to address speaker identification. Conversely, the standard CNN does not exhibit such a meaningful pattern: the CNN filters tend to correctly focus on the lower part of the spectrum, but peaks tuned on first and second formants do not clearly appear. As one can observe from Fig. \ref{fig:cum}, the CNN curve stands above the SincNet one. SincNet, in fact, learns filters that are, on average, more selective than CNN ones, possibly better capturing narrow-band speaker clues.

Fig. \ref{fig:asr_cum} shows the cumulative frequency response of a CNN and SincNet obtained on a noisy speech recognition task. In this experiment, we have artificially corrupted TIMIT with a significant quantity of noise in the band between 2.0 and 2.5 kHz (see the spectrogram) and we have analyzed how fast the two architectures learn to avoid such a useless band. The second row of sub-figures compares the CNN and the SincNet at a very early training stage (i.e., after having processed only one hour of speech in the first epoch), while the last row shows the cumulative frequency responses after completing the training. From the figures emerges that both CNN and SincNet have correctly learned to avoid the corrupted band at end of training, as highlighted by the holes between 2.0 and 2.5 kHz in the cumulative frequency responses. SincNet, however, learns to avoid such a noisy band much earlier. In the second row of sub-figures, in fact, SincNet shows a visible valley in the cumulative spectrum even after processing only one hour of speech, while CNN has only learned to give more importance to the lower part of the spectrum.

\end{itemize}

\section{Related Work} \label{sec:rel_work}
Several works have recently explored the use of low-level speech representations to process audio and speech with CNNs. Most prior attempts exploit magnitude spectrogram features \cite{e2e_spk_id,spk_rec_time_freq,voxceleb,tara_asru2013,learn_fbank_const,fbank_par}. Although spectrograms retain more information than standard hand-crafted features, their design still requires careful tuning of some crucial hyper-parameters, such as the duration, overlap, and typology of the frame window, as well as the number of frequency bins. For this reason, a more recent trend is to directly learn from raw waveforms, thus completely avoiding any feature extraction step. 
This approach has shown promise in speech \cite{palaz_raw,tara_raw,google_rawmulti,joint7,tuske}, including emotion tasks \cite{dnn_emotion}, speaker recognition \cite{raw_speaker_id}, spoofing detection \cite{spoofing_raw}, and speech synthesis \cite{wavenet,sample_rnn}.

Similar to SincNet, some previous works have proposed to add constraints on the CNN filters, for instance forcing them to work on specific bands \cite{tara_asru2013,learn_fbank_const}.  
Differently from the proposed approach, the latter works operate on spectrogram features and still learn all the L elements of the CNN filters. An idea related to the proposed method has been recently explored in \cite{fbank_par}, where a set of parameterized Gaussian filters are employed. This approach operates on the spectrogram domain, while SincNet directly considers the raw waveform in the time domain. Similarly to our work, in \cite{Neil_ICASSP2018} the convolutional filters are initialized with a predefined filter shape.  However, rather than focusing on cut-off frequencies only,  all the basic taps of the FIR filters are still learned.

Some valuable works have recently proposed theoretical and experimental frameworks to analyze CNNs \cite{CNN_theory,CNN_mallat}.  In particular, \cite{Palaz_CNN_an,raw_speaker_id,Magimai_CNN_an} feed a standard CNN with raw audio samples and analyze the filters learned in the first layer on both speech recognition and speaker identification tasks. The authors highlight some interesting properties emerged from analyzing the cumulative frequency response and propose a spectral dictionary interpretation of the learned filters. Similarly to our findings, the latter works noticed that the filters tend to focus more on the lower part of the spectrum and they can sometimes highlight some peaks that likely corresponds to the fundamental frequency. In this work, we argue that all of these interesting properties can be observed more clearly and at an earlier training stage with SincNet.

This paper extends our previous studies on the SincNet \cite{SincNet}. To the best of our knowledge,  this paper is the first that shows the effectiveness of the proposed SincNet in a speech recognition application. Moreover, this work not only considers standard close-talking speech recognition, but it also extends the validation of SincNet to distant-talking speech recognition \cite{ravanelli_thesis,ravanelli_icassp,ravanelli15}. 


\section{Results} \label{sec:exp}
The proposed SincNet has been evaluated on both speech and speaker recognition using different corpora. This work considers a challenging but realistic speaker recognition scenario: for all the adopted corpora, we only employed 12-15 seconds of training material for each speaker, and we tested the system performance on short sentences lasting from 2 to 6 seconds.
In the spirit of reproducible research, we release the code of SincNet for speaker identification\footnote{\label{foot:code} at \url{https://github.com/mravanelli/SincNet/}.} and speech recognition\footnote{\label{foot:code} at \url{https://github.com/mravanelli/pytorch-kaldi/}.} (under the PyTorch-Kaldi project \cite{pytorch_kaldi}). More details on the adopted datasets as well as on the SincNet and baseline setups can found in the \textbf{appendix}.

\subsection{Speaker Recognition}

Table \ref{tab:spk_id_res} reports the Classification Error Rates (CER\%) achieved on a speaker-id task. The table shows that  SincNet outperforms other systems on both TIMIT (462 speakers) and Librispeech (2484 speakers) datasets. The gap with a standard CNN fed by raw waveform is larger on TIMIT, confirming the effectiveness of SincNet when few training data are available. Although this gap is reduced when LibriSpeech is used, we still observe a 4\% relative improvement that is also obtained with faster convergence (1200 vs 1800 epochs). 
Standard FBANKs provide results comparable to SincNet only on TIMIT, but are significantly worse than our architecture when using Librispech. With few training data, the network cannot discover filters that are much better than that of FBANKs, but with more data a customized filter-bank is learned and exploited to improve the performance.

 Table \ref{tab:spk_ver_res} extends our validation to speaker verification, reporting the Equal Error Rate (EER\%) achieved with Librispeech. All DNN models show promising performance, leading to an EER lower than 1\% in all cases. The table also highlights that SincNet outperforms the other models, showing a relative performance improvement of about 11\% over the standard CNN model. Note that the speaker verification system is derived from the speaker-id neural network using the \textit{d-vector} technique. The \textit{d-vector} \cite{dnn_spk_rec_class2,voxceleb} is extracted from the last hidden layer of the speaker-id network. A speaker-dependent d-vector is computed and stored for each enrollment speaker by performing an L2 normalization and averaging all the d-vectors of the different speech chunks. The cosine  distance between enrolment and test d-vectors is then calculated, and a threshold is then applied on it to reject or accept the speaker.
Ten utterances from impostors were randomly selected for each sentence coming from a genuine speaker.
To assess our approach on a standard open-set speaker verification task, all the enrolment and test utterances were taken from a speaker pool different from that used for training the speaker-id DNN.
 
    \begin{table}[t!]
        \begin{minipage}{0.55\textwidth}
            \centering
               \begin{tabular}{llr}  
                \toprule
                & TIMIT & LibriSpeech  \\ 
                \midrule
                DNN-MFCC             &   0.99      &  2.02      \\ 
                CNN-FBANK       &   0.86      &  1.55     \\ 
                CNN-Raw       &   1.65     &  1.00       \\ 
                SincNet       &   \textbf{0.85}      &  \textbf{0.96}      \\ 
                \bottomrule
            \end{tabular}
            \caption{Classification Error Rate (CER\%) of speaker identification systems trained on TIMIT (462 spks) and Librispeech (2484 spks) datasets. SincNet outperforms the
competing alternatives.}
             \label{tab:spk_id_res}

        \end{minipage}
        \hfillx
        \begin{minipage}{0.4\textwidth}
            \centering
            \begin{tabular}{lr}  
            \toprule
            & EER(\%)  \\ 
            \midrule
            DNN-MFCC   &  0.88  \\ 
            CNN-FBANK   &  0.60  \\ 
            CNN-Raw     &  0.58 \\ 
            SINCNET     &  \textbf{0.51}    \\
            \bottomrule
            \end{tabular}
            \caption{Speaker Verification Equal Error Rate (EER\%) on Librispeech datasets over different systems. SincNet outperforms the competing alternatives.}
            \label{tab:spk_ver_res}
        \end{minipage}
    \end{table}

For the sake of completeness, experiments have also been conducted with standard i-vectors.
Although a detailed comparison with this technology is out of the scope of this paper, it is worth noting that our best i-vector system achieves an EER=1.1\%, rather far from what is achieved with DNN systems. It is well-known in the literature that i-vectors provide competitive performance when more training material is used for each speaker and when longer test sentences are employed \cite{i-vect_short,i-vect_short2,i-vect_short3}. Under the challenging conditions faced in this work, neural networks achieve better generalization.

\subsection{Speech Recognition}
Tab. \ref{tab:asr} reports the speech recognition performance obtained by CNN and SincNet using the TIMIT and the DIRHA dataset \cite{dirha_asru}. To ensure a more accurate comparison between the architectures, five experiments varying the initialization seeds were conducted for each 
model and corpus. Table \ref{tab:asr} thus reports the average speech recognition performance. Standard deviations, not reported here, range between $0.15$ and $0.2$ for all the experiments.

\begin{table}[h]

\centering

\begin{tabular}{llr}  
\toprule
& TIMIT & DIRHA  \\ 
\midrule
CNN-FBANK       &   18.3      &  40.1     \\ 
CNN-Raw waveform       &   18.1     &  40.0       \\ 
SincNet-Raw waveform      &   \textbf{17.2}      &  \textbf{37.2}      \\ 
\bottomrule
\end{tabular}
\caption{Speech recognition performance obtained on the TIMIT and DIRHA datasets.}
\label{tab:asr}
\end{table}

For all the datasets, SincNet outperforms CNNs trained on both standard FBANK and raw waveforms. The latter result confirms the effectiveness of SincNet not only in close-talking scenarios but also in challenging noisy conditions characterized by the presence of both noise and reverberation. As emerged in Sec.\ref{sec:sinc},  SincNet is able to effectively tune its filter-bank front-end to better address the characteristics of the noise.

\section{Conclusions and Future Work}
\label{sec:conc}
This paper proposed SincNet, a neural architecture for directly processing waveform audio. Our model, inspired by the way filtering is conducted in digital signal processing, imposes constraints on the filter shapes through efficient parameterization. SincNet has been extensively evaluated on challenging speaker and speech recognition tasks, consistently showing some performance benefits. 

Beyond performance improvements, SincNet also significantly improves convergence speed over a standard CNN, is more computationally efficient due to the exploitation of filter symmetry, and it is more interpretable than standard black-box models. Analysis of the SincNet filters, in fact, revealed that the learned filter-bank is tuned to the specific task addressed by the neural network.
In future work, we would like to evaluate SincNet on other popular speaker recognition tasks, such as VoxCeleb. Inspired by the promising results obtained in this paper, in the future we will explore the use of SincNet for supervised and unsupervised speaker/environmental adaptation. 
Moreover, although this study targeted speaker and speech recognition only, we believe that the proposed approach defines a general paradigm to process time-series and can be applied in numerous other fields. 

\section*{Acknowledgement}
This research was enabled in part by support provided by Calcul Qu\'ebec and Compute Canada.

\medskip

\small

\bibliographystyle{unsrt}
\bibliography{ref.bib}

\newpage

\appendix
\section*{Appendix}
\subsection*{Corpora}
To provide experimental evidence on datasets characterized by different numbers of speakers, this paper considers the TIMIT (462 spks, \textit{train} chunk)  \cite{timit} and Librispeech  (2484 spks) \cite{librispeech} corpora. For speaker verification experiments, non-speech intervals at the beginning and end of each sentence were removed. Moreover, the Librispeech sentences with internal silences lasting more than 125 ms were split into multiple chunks. To address text-independent speaker recognition, the calibration sentences of TIMIT (i.e., the utterances with the same text for all speakers) have been removed. For the latter dataset, five sentences for each speaker were used for training, while the remaining three were used for test. For the Librispeech corpus, the training and test material have been randomly selected to exploit 12-15 seconds of training material for each speaker and test sentences lasting 2-6 seconds. To evaluate the performance in a challenging distant-talking scenario, speech recognition experiments have also considered the DIRHA dataset \cite{dirha_asru}. This corpus, similarly to the other DIRHA corpora \cite{dirha_grid,dirha_icassp}, has been developed in the context of the DIRHA project \cite{lrec} and is based on WSJ sentences \cite{wsj_corpus} recorded in a domestic environment. Training is based on contaminating WSJ-5k utterances with realistic impulse responses \cite{Ravanelli-12,rav_in14}, while the test phase test phase consists of 409 WSJ sentences recorded by native speakers in a domestic environment (the average SNR is 10 dB).

\subsection*{SincNet Setup}
The waveform of each speech sentence was split into chunks of 200 ms (with 10 ms overlap), which were fed into the SincNet architecture. The first layer performs sinc-based convolutions as described in Sec. \ref{sec:sinc}, using 80 filters of length $L=251$ samples. The architecture then employs two standard convolutional layers, both using 60 filters of length 5. Layer normalization \cite{layer_norm} was used for both the input samples and for all convolutional layers (including the SincNet input layer). Next, three fully-connected layers composed of 2048 neurons and normalized with batch normalization \cite{batchnorm,ravanelli_SLT} were applied. All hidden layers use leaky-ReLU \cite{leaky_relu} non-linearities. The parameters of the sinc-layer were initialized using mel-scale cutoff frequencies, while the rest of the network was initialized with the well-known ``Glorot" initialization scheme \cite{xavier}. Frame-level speaker and phoneme classifications were obtained by applying a softmax classifier, providing a set of posterior probabilities over the targets. For speaker-id, a sentence-level classification was simply derived by averaging the frame predictions and voting for the speaker which maximizes the average posterior.
Training used the RMSprop optimizer, with a learning rate $lr=0.001$, $\alpha=0.95$, $\epsilon=10^-7$, and minibatches of size 128. 
All the hyper-parameters of the architecture were tuned on TIMIT, then inherited for Librispeech as well. 
The speaker verification system was derived from the speaker-id neural network using the \textit{d-vector} approach \cite{dnn_spk_rec_class2,voxceleb}, which relies on the output of the last hidden layer and computes the cosine  distance between test and the claimed speaker d-vectors. 
Ten utterances from impostors were randomly selected for each sentence coming from a genuine speaker. 
Note that to assess our approach on a standard open-set speaker-id task, all the impostors were taken from a speaker pool different from that used for training the speaker-id DNN.

\subsection*{Baseline Setups}
We compared SincNet with several alternative systems. 
First, we considered a standard CNN fed by the raw waveform. This network is based on the same architecture as SincNet, but replacing the sinc-based convolution with a standard one. 

A comparison with popular hand-crafted features was also performed. To this end, we computed 39 MFCCs (13 static+$\Delta$+$\Delta\Delta$) and 40 FBANKs using the Kaldi toolkit \cite{kaldi_short}. These features,  computed every 25 ms with 10 ms overlap, were gathered to form a context window of approximately 200 ms (i.e., a context similar to that of the considered waveform-based neural network). A CNN was used for FBANK features, while a Multi-Layer Perceptron (MLP) was used for MFCCs. Note that CNNs exploit local correlation across features and cannot be effectively used with uncorrelated MFCC features. Layer normalization was used for the FBANK network, while batch normalization was employed for the MFCC one. The hyper-parameters of these networks were also tuned using the aforementioned approach. 

For speaker verification experiments, we also considered an i-vector baseline. The i-vector system was implemented with the SIDEKIT toolkit \cite{sidekit}. The GMM-UBM model, the Total Variability (TV) matrix, and the Probabilistic Linear Discriminant Analysis (PLDA) were trained on the Librispeech data (avoiding test and enrollment sentences). GMM-UBM was composed of 2048 Gaussians, and the rank of the TV and PLDA eigenvoice matrix was 400. The enrollment and test phase is conducted on Librispeech using the same set of speech segments used for DNN experiments.

\end{document}